\documentclass{elsart}
\usepackage[square,sort&compress,comma]{natbib}
\usepackage[dvips]{graphicx}
\usepackage{amsmath}
\usepackage{amsfonts,amssymb,latexsym} 

\usepackage[english]{babel} 

\begin{document}








 

 
 
\newlength{\twocolumnwidth}\setlength{\twocolumnwidth}{246pt} 
 
 
 
\begin{frontmatter}



\title{Quantum Dynamics of Atomic Coherence\\ in a Spin-1 Condensate:\\ 
Mean-Field versus Many-Body Simulation}


\author{L.I.\ Plimak,}
\author{C.\ Wei\ss,}
\author{R.\ Walser, and}
\author{W.\ P.\ Schleich}
\address{Abteilung Quantenphysik, Universit\"at Ulm, 
D-89069 Ulm, Germany .}

\begin{abstract}
We analyse and numerically simulate  
the full many-body quantum dynamics of a spin-1 condensate 
in the single spatial mode approximation. 
Initially, the condensate is  
in a ``ferromagnetic'' 
state with all spins aligned along the $y$ axis and the magnetic field 
pointing along the $z$ axis. 
In the course of evolution the spinor condensate 
undergoes a characteristic change of symmetry, which in a real experiment 
could be a signature 
of spin-mixing many-body interactions. 
The results of our simulations are 
conveniently visualised within 
the picture of irreducible tensor operators. 
\end{abstract}


\end{frontmatter}
\section{Introduction}
This paper is a tribute to Bruce Shore's role in establishing the 
concept of atomic coherence and the related techniques of irreducible tensor 
operators \cite{Shore1989}. 
We demonstate that this viewpoint  
emerges quite naturally when describing the evolution of a spinor 
condensate. 
During the past decade 
condensed Bose gases have energed as 
flexible test system to explore the
rich structure of many-body physics. 
A systematic introduction into cold gases in 
general and spinor condensates in particular may be found, e.g., 
in the book 
by Pethick and Smith \cite{Pethick2002}.
With the development of purely optical traps \cite{AllOptTrap,Chapman2001,Sengstock2004},
we are in the position to hold atoms with their complete hyperfine
submanifold, thus entering the regime of spinor physics 
\cite{Pethick2002,Ho1998,Machida1998,Bigelow1998,Ketterle1999}. 
This is exactly the condition under which one expects the techniques 
of atomic coherence \cite{Shore1989,Omont1977,Varshalovich1988} 
to be most helpful. 

\begin{figure}[h]
 \begin{center} 
 \includegraphics[width=0.5\textwidth]{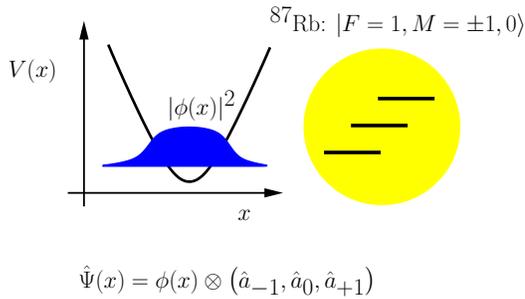}
 \caption{Schematic diagram of the harmonically 
 trapped spin-1 $^{87}$Rb condensate in the single-mode 
	 approximation.}
\label{Schema}
\end{center}
\end{figure}

There are two major approximations used in the literature 
when describing spinor condensates theoretically: the common 
mean-field approximation 
(MF)
and the single mode aproximation (SMA). 
Within the SMA, the spatial dependence of the Bose field operator 
describing the spinor gas is assumed given and decoupled from the 
internal dynamics, 
as illustrated schematically in Fig.~\ref{Schema}. 
The reason one
can separate the local redistribution of particles amongst hyperfine sublevels
due to collisions from the long range dynamics is clear:
a typical trapped
alkali atom is paramagnetic, but provides only a single Bohr magneton of
magnetic moment. 
Consequently the resulting magnetic dipolar interaction is very weak
compared to the typical interactomic Van-der-Waals energy. 

The 
single mode approximation 
is commonly introduced on top of the 
mean-field approximation. 
One exception is  the paper by Koashi and Ueda 
\cite{Koashi2000} 
where the SMA is introduced directly into the many-body 
treatment 
(see also \cite{Pethick2002}).  
This, and also neglecting the quadratic Zeeman part of the 
energy, allowed the authors to construct an explicit solution to the 
problem including collisions. 
In this paper we 
include the quadratic Zeeman 
part which plays a crucial role in the dynamics of the condensate 
as realized with $^{87}$Rb or $^{23}$Na atoms 
\cite{Sengstock2004,Chapman2005,Sengstock2005}. 
Consequently, we are left with an interacting 
three mode Hamiltonian, whose dynamics will be the subject of study in this
article.


It should be pointed out that our approach would not apply 
directly to the recently 
realised Chromium Bose-Einstein condensate 
\cite{Pfau2005} where the dipolar interaction is comparable to 
the Van-der-Waals energy.   
Even less so in the case of heteronuclear molecules pursued by many 
experimental groups around the world where the dipolar interaction is 
expected to be the dominant mechanism. 
Independence of the internal and spatial motion is completely 
lifted e.g. in the proposed 
Einstein--de Haas effect \cite{Pfau2006,Ueda2005}. 
The SMA is obviously not aplicable under the conditions for condensate 
fragmentation and domain formation 
(see \cite{You2005-2} and references therein).

The article is organized as follows: In Section~\ref{SecManyBody} we 
review the derivation
of the three mode spinor Hamiltonian. For later reference, we will also
discuss the standard mean-field approximation in the language of Hamiltonian
mechanics in Section~\ref{SecMF}. 
Visualizing the results of a full many-body evolution of
the three mode system is a nontrivial problem. 
Therefore, we introduce the
concepts of irreducible tensor operators in Section~\ref{SecCohEv}. 
Finally, we discuss and
compare the results of the many-body calculation in Section~\ref{SecResDisc}.
\section{Many-body approach to spinor condensate dynamics}
\label{SecManyBody}
The $F=1$ bosons are described in the usual way by 
a field operator with the three spherical components, 
$\hat \psi _k({\mbox{\rm\boldmath$r$}})$, where $k = 1,0,-1$ (in this order) 
labels the $z$ projection of the atomic spin.
The effective low energy Hamiltonian of such bosons in a homogeneous 
magnetic field is written as: 
(cf.~\cite{Sengstock2004,Ho1998,Chapman2005,You2005}) 
\begin{align} 
\begin{aligned}
\hat{\mathcal H} = 
\hat{\mathcal H}_{0} + 
\hat{\mathcal H} _{\text{B}} + 
\hat{\mathcal H}^{(0)}_{\text{coll}} 
+ 
\hat{\mathcal H}^{(2)}_{\text{coll}} 
\, . 
\end{aligned}%
\end{align}%
Here, 
\begin{align} 
\begin{aligned}
\hat{\mathcal H}_{0} & = \sum_k
\int d^3 {{\mbox{\rm\boldmath$r$}}}\hat\psi^{\dag} _k({\mbox{\rm\boldmath$r$}})
\Big[
-\frac{\hbar^2}{2M} 
\Delta 
+ U({\mbox{\rm\boldmath$r$}})
\Big]
\hat\psi_k({\mbox{\rm\boldmath$r$}})
\end{aligned}%
\end{align}%
contains the kinetic energy and the trapping potential which holds the atoms 
in place, 
and 
\begin{align} 
\begin{aligned}
\hat{\mathcal H}_{\text{B}} & = \sum_k E_k
\int d^3 {{\mbox{\rm\boldmath$r$}}}\hat\psi^{\dag} _k({\mbox{\rm\boldmath$r$}})
\hat\psi_k({\mbox{\rm\boldmath$r$}}) = \sum_k E_k \hat{\mathcal N}_k
\end{aligned}%
\end{align}%
describes the effects of the magnetic field. 
For weaker magnetic fields, this is mainly the 
linear Zeeman effect, 
$E_{1}-E_{0}\approx E_{0}-E_{-1}\propto B$, and 
the quadratic 
Zeeman, or Paschen-Back, effect, 
$E_{1}+E_{-1}- 2 E_{0}\propto B^2$. 
The latter is crucial for the purposes of our analyses.  
The general Breit-Rabi formulas for $E_k$ may be found in Ref.\ \cite{Vanier1988}.

Both $\hat{\mathcal H}_{0}$ and $\hat{\mathcal H} _{\text{B}}$ describe 
single-body physics.  
Two-body collisions enter via 
$\hat{\mathcal H}^{(0)}_{\text{coll}}$ and $\hat{\mathcal H}^{(2)}_{\text{coll}}$. 
The former describes the familiar spin-independent density-density 
interaction, 
\begin{align} 
\begin{aligned}
\hat{\mathcal H}^{(0)}_{\text{coll}} = 
\frac{c_0}{2} \sum_{k,l} \int d^3{\mbox{\rm\boldmath$r$}}
\hat\psi^{\dag} _k({\mbox{\rm\boldmath$r$}})
\hat\psi^{\dag} _l({\mbox{\rm\boldmath$r$}})
\hat\psi _l({\mbox{\rm\boldmath$r$}}) 
\hat\psi _k({\mbox{\rm\boldmath$r$}})
= \frac{c_0}{2} \int d^3{\mbox{\rm\boldmath$r$}}
\,{\mbox{\rm\boldmath$:$}}\,\hat n^2 ({\mbox{\rm\boldmath$r$}})
\,{\mbox{\rm\boldmath$:$}}\, , 
\end{aligned}%
\end{align}%
where $\hat n ({\mbox{\rm\boldmath$r$}}) = \sum_k\hat n_k ({\mbox{\rm\boldmath$r$}}) = \sum_k\hat\psi^{\dag} _k({\mbox{\rm\boldmath$r$}})
\hat\psi _k({\mbox{\rm\boldmath$r$}})$, and ${\mbox{\rm\boldmath$:$}}{\mbox{\rm\boldmath$:$}}$ stands for the normal ordering of the field operators. 
The latter describes the spin-dependent part of the collision, 
\begin{align} 
\begin{aligned}
\hat{\mathcal H}^{(2)}_{\text{coll}} = 
\frac{c_2}{2}\int d^3{\mbox{\rm\boldmath$r$}}
\,{\mbox{\rm\boldmath$:$}}\,\hat {{\mbox{\rm\boldmath$\mathcal F$}}}^2 ({\mbox{\rm\boldmath$r$}})
\,{\mbox{\rm\boldmath$:$}}\, , 
\end{aligned}%
\end{align}%
where $\hat {{\mbox{\rm\boldmath$\mathcal F$}}} ({\mbox{\rm\boldmath$r$}})$ is the density of the angular momentum of 
the condensate, 
\begin{align} 
\begin{aligned}
\hat {{\mathcal F}_s}({\mbox{\rm\boldmath$r$}}) = \sum_{k,l}[F_s]_{kl}\hat\psi^{\dag} _k({\mbox{\rm\boldmath$r$}})
\hat\psi _l({\mbox{\rm\boldmath$r$}}) . 
\end{aligned}%
\end{align}%
The Cartesian components 
of the single-particle spin operator, ${\mbox{\rm\boldmath$F$}}$, are 
defined as: \cite{Varshalovich1988}
\begin{gather}
\begin{gathered}
F_x=\frac{1}{\sqrt{2}}
\begin{pmatrix}
 0&1&0\\
 1&0&1\\
 0&1&0
 \end{pmatrix}, \ \ 
 F_y=\frac{i}{\sqrt{2}}
\begin{pmatrix}
 0&-1&0\\
 1&0&-1\\
 0&1&0
 \end{pmatrix},\ \ 
F_z=
\begin{pmatrix}
 1&0&0\\
 0&0&0\\
 0&0&-1
 \end{pmatrix}.
\end{gathered}
\end{gather}
The collisional interaction constants, $c_0$ and $c_2$, are 
expressed in terms of the scattering lengths $a_0$ and $a_2$, 
for two spin-1 atoms in the combined symmetric channels of total spins, 
respectively, 0 and 2, as \cite{Ho1998}
\begin{align} 
\begin{aligned}
c_0 &= \frac{4\pi \hbar^2(a_0+2a_2)}{3M} , 
\ \
c_2 &= \frac{4\pi \hbar^2(a_2-a_0)}{3M} 
. 
\end{aligned}%
\end{align}%

A typical value of $c_2$ is one-two orders of magnitude below $c_0$. 
This leads to a natural separation of time scales: 
on smaller time scales, one may neglect $\hat{\mathcal H}^{(2)}_{\text{coll}}$, 
whereas for longer time scales it may be of crucial importance. 
The ultimate example of a longer time scale is the 
ground state of the condensate, which may be polar or ferromagnetic 
depending on the sign of $c_2$ \cite{Ho1998}. 
An example of the opposite are experiments where the spinor condensate is prepared in a particular magnetic state, and the evolution of the 
condensate is then observed \cite{Sengstock2004,Chapman2005,Chapman2004}. 
In this case, an approximation commonly used in the literature is the so-called 
single mode approximation (SMA). 
Technically, SMA is recovered by assuming that the spatial dependence 
of the field operator is fixed, $\hat\psi_k({\mbox{\rm\boldmath$r$}}) = \phi({\mbox{\rm\boldmath$r$}})\hat a_k
$, $\int d^3 {\mbox{\rm\boldmath$r$}} |\phi({\mbox{\rm\boldmath$r$}})|^2 = 1$. 
Under the SMA, the Hamiltonian of the 
system of $N$ spin-one particles becomes: (cf.\ \cite{Koashi2000})
\begin{multline} 
\hspace{0.4\columnwidth}\hspace{-98.4pt}
\hat {\mathcal H} = -p \widehat{F_z} + q 
 \widehat{F^2_z} 
+
 \frac{g_2}{2} 
\mbox{$:\hat{{\mbox{\rm\boldmath${\mathcal F}$}}}^2\!\! :$} 
\\ = 
-p \hat{\mathcal F}_z + q 
 \widehat{F^2_z} 
+ \frac{g_2}{2} 
\left(
 \hat{{\mbox{\rm\boldmath${\mathcal F}$}}}^2 - 2 \hat{\mathcal N}
\right)
. 
\label{eq:HFockSpace} 
\hspace{0.4\columnwidth}\hspace{-98.4pt}
\end{multline}%
Here, 
$g_2 = c_2 \int d^3 {\mbox{\rm\boldmath$r$}} |\phi({\mbox{\rm\boldmath$r$}})|^4$, 
$p = (E_{-1}-E_{1})/2$, and $q = (E_{1}+E_{-1})/2-E_0$. 
The ``wide hat'' symbol denotes mapping of a single-body 
operator into the Fock space, 
\begin{align} 
\begin{aligned}
\widehat{A} = \sum_{k,l} \hat a_k^{\dag} A_{kl}\hat a_l, 
\end{aligned}%
\label{eq:ToFock} 
\end{align}%
with $\hat a_l$ being the mode annihilation operators; in our case, 
$l=1,0,-1$ (in this order). 
The ``wide hat'' should not be confused with the 
``normal'' hat which just signifies that a 
quantity is an operator in the Fock space. 
In particular, 
$\hat {{\mbox{\rm\boldmath${\mathcal F}$}}}
 = \widehat{{\mbox{\rm\boldmath$F$}}}$ is the full angular momentum operator, 
and $\hat {\mathcal N}= \widehat{{\mathcal I}_3}$ is the Fock-space operators 
of the total particle number, where 
${\mathcal I}_3$ is a 3$\times$3 unit matrix.  
The second line in (\ref{eq:HFockSpace}) is found using that, 
\begin{align} 
\begin{aligned}
\mbox{$:\hat{{\mbox{\rm\boldmath${\mathcal F}$}}}^2\!\! :$} = \hat{{\mbox{\rm\boldmath${\mathcal F}$}}}^2 - \widehat{{\mbox{\rm\boldmath$F$}}^2} = 
\hat{{\mbox{\rm\boldmath${\mathcal F}$}}}^2 - 2 \hat{\mathcal N} 
. 
\end{aligned}%
\end{align}%

In numerical terms, constituents of (\ref{eq:HFockSpace}) are sparse 
matrices; these are handled very efficiently by common 
high-level numerical 
packages. 
Some details on how these were implemented 
numerically may be found in Appendix \ref{ToFock}. 
\section{The canonical formulation of the mean-field dynamics}
\label{SecMF}
\subsection{Introducing the mean field}
In the mean field approximation, operators $\hat{a}_k$ are 
replaced by complex numbers $\zeta_k$. 
However, this transition is not uniquely defined. 
Using different orderings (normal, symmetric, etc.) of creation and annihilation operator yields different {\em functional forms\/} of the 
quantum Hamiltonian. 
In quantum mechanics these functional forms represent the 
same Hamiltonian. 
In the mean field picture where 
ordering is immaterial for the complex numbers $\zeta_k$, $\zeta_k^\ast$, 
these different functional forms result in different classical 
Hamiltonians. 
In the scalar case, the corresponding corrections are of the order of $1/N$, 
where $N$ is the total number of atoms. 
For the spinor condensate, these corrections are of the order of one over 
the population of Zeeman sublevels. 
There can well be pathological cases when one level is much depleted, 
leading to large corrections. 
It is therefore of importance 
to have the procedure of introducing the mean field approximation 
for the spinor condensate clearly defined. 

There is a number of ways the mean field picture can be introduced. 
The most rigorous one is via the phase-space technique 
(see, e.g., \cite{Steel} and references therein), where the time-dependent 
Gross-Pitaevskii equation appears on dropping quantum noises in the equations 
for the phase-space amplitudes. 
This procedure is equivalent to using the normally-ordered form 
of the interaction Hamiltonian, replacing the field operators by the 
c-numbers, and then {\em postulating\/} the resulting c-number 
function to be the classical Hamiltonian. 
The lack of rigour is clearly seen in the fact that this procedure 
leaves it unclear how quantum averages are expressed by the c-number 
amplitudes. 
Some insight can be gained from the fact that if we postulate the 
many body state as a coherent state, 
$\left|{\mbox{\rm\boldmath$\zeta$}}\right\rangle=\left|\zeta_1\right\rangle\left|\zeta_0\right\rangle\left|\zeta_{-1}\right\rangle$, 
and then minimize the energy subject to the condition 
$\left\langle {\mbox{\rm\boldmath$\zeta$}}\left|\hat{\mathcal N}\right|{\mbox{\rm\boldmath$\zeta$}}\right\rangle = ||{\mbox{\rm\boldmath$\zeta$}}||^2 = N$, 
the time-independent Gross-Pitaevskii equation is recovered. 
This recipe is in fact general: 
a correct mean-field expression for the average of an operator $\hat{\mathcal X}$ 
is found as $\left\langle \hat{\mathcal X}\right\rangle = 
\left\langle {\mbox{\rm\boldmath$\zeta$}}\left|\hat{\mathcal X}\right|{\mbox{\rm\boldmath$\zeta$}}\right\rangle$. 
This is equivalent to replacing field operators by c-numbers in 
the {\em normally ordered representation\/} of $\hat{\mathcal X}$. 

\subsection{The classical Hamilton equations of motion}
Following the above recipe, we postulate the classical 
Hamilton function to be, 
\begin{align} 
\begin{aligned}
H\!\left ({\mbox{\rm\boldmath$\zeta$}}, 
{\mbox{\rm\boldmath$\pi$}}\right ) &= 
\left\langle {\mbox{\rm\boldmath$\zeta$}}\left|\hat{\mathcal H}\right|{\mbox{\rm\boldmath$\zeta$}}\right\rangle 
\\ 
&= 
-p\big(|\zeta_{1}|^2-|\zeta_{-1}|^2\big)+q\big(|\zeta_{1}|^2+|\zeta_{-1}|^2\big)
\\
&\hspace{1.5em}+\frac{g_2}{2}\Big [
|\zeta_{1}|^4+|\zeta_{-1}|^4-2|\zeta_{1}|^2|\zeta_{-1}|^2+2|\zeta_0|^2|\zeta_{1}|^2
\\ &\hspace{1.5em}+2|\zeta_0|^2|\zeta_{-1}|^2+2\zeta_0^2\zeta_{1}^\ast\zeta_{-1}^\ast+2{\zeta_0^\ast}^2\zeta_{1}\zeta_{-1}\Big], 
\end{aligned}%
\label{eq:HamiltonFunction} 
\end{align}%
where (\ref{eq:HFockSpace}) was used to find the explicit form of 
the Hamilton function.  
To use the powerfull methods of analytical mechanics 
we have to know the canonical conjugated momentum ${\mbox{\rm\boldmath$\pi$}}$ to the coordinate ${\mbox{\rm\boldmath$\zeta$}}$. If, by definition, $\mathbf{x}$ and $\mathbf{p}$ are three-dimensional canonical conjugated variables, defining the complex variable ${\mbox{\rm\boldmath$\zeta$}}$
\begin{equation}
{\mbox{\rm\boldmath$\zeta$}}=\frac{\mathbf{x}+i\mathbf{p}}{\sqrt{2}},
\end{equation} 
than we can identify ${\mbox{\rm\boldmath$\pi$}}=i{\mbox{\rm\boldmath$\zeta$}}^\ast$. One can verify easily, that this is a proper canonical transform, as
\begin{equation}
\left \{\zeta_i,\pi_l\right \}_{\left \{x,p\right \}}=\delta_{il},
\end{equation} 
with the following conventional definition 
of the Poisson brackets,  
\begin{equation}
\left \{F,G\right\}_{\left \{x,p\right \}}=\sum_{i=\pm 1,0}\frac{\partial F}{\partial x_i}\frac{\partial G}{\partial p_i}-\frac{\partial F}{\partial p_i}\frac{\partial G}{\partial x_i} . 
\end{equation} 
So we obtain the Hamilton equations of motion 
\begin{equation}
\begin{split}
\frac{d}{dt}\zeta_l&=\left \{\zeta_l,H\!\left ({\mbox{\rm\boldmath$\zeta$}},{\mbox{\rm\boldmath$\pi$}}\right)\right\}_{\left \{\zeta,\pi\right \}}=\frac{\partial H\!\left ({\mbox{\rm\boldmath$\zeta$}},{\mbox{\rm\boldmath$\pi$}}\right)}{\partial\pi_l},\\
\frac{d}{dt}\pi_l&=\left \{\pi_l,H\!\left ({\mbox{\rm\boldmath$\zeta$}},{\mbox{\rm\boldmath$\pi$}}\right)\right\}_{\left \{\zeta,\pi\right \}}=-\frac{\partial H\!\left ({\mbox{\rm\boldmath$\zeta$}},{\mbox{\rm\boldmath$\pi$}}\right)}{\partial\zeta_l}.
\end{split}
\end{equation}
By employing complex coordinates we have doubled the dimension of the phase 
space making the two sets of equations of motion redundant. Explicitly, we find the three components of the 
time-dependent Gross-Pitaevskii equation
\begin{gather}
\label{eq:GP-Equation}
i\frac{d}{dt}{\mbox{\rm\boldmath$\zeta$}}=\left (H_0+g_2H_1\right )
{\mbox{\rm\boldmath$\zeta$}},\\
H_0=\left [\begin{matrix}-p+q&0&0\\0&0&0\\0&0&p+q\end{matrix}\right ],\ \ 
H_1=\left [\begin{matrix}|\zeta_0|^2+\mathcal{F}_z & \zeta_{-1}^\ast\zeta_0 & 0\\ \zeta_{-1}\zeta_0^\ast &|\zeta_{1}|^2+|\zeta_{-1}|^2 &\zeta_{1}\zeta_0^\ast\\0 & \zeta_{1}^\ast\zeta_0 & |\zeta_0|^2-\mathcal{F}_z\end{matrix}\right ],
\end{gather}
with $\mathcal{F}_z=|\zeta_{1}|^2-|\zeta_{-1}|^2$. 
For the initial condition as in the experiment of the Sengstock 
group \cite{Sengstock2004}, these nonlinear equations have been solved analytically in terms of the periodic Jacobi elliptic functions, see \cite{Sengstock2005}. 
\subsection{Comparing the many-body and mean-field dynamics}
\label{EigSVD}
Consider now how an effective mean-field amplitude is recovered 
in the many-body techniques. Were the population and coherence 
properties of the matter field investigated in a measurement, all the 
relevant information enters through the single-body density matrix, 
\begin{align} 
\begin{aligned}
\rho _{lk}(t) = \left\langle \Psi_0\left|\hat a^{\dag}_k(t) \hat a _l(t)\right|\Psi_0\right\rangle, 
\end{aligned}%
\label{eq:rhott} 
\end{align}%
where $\left|\Psi_0\right\rangle$ is the initial many-body state, 
and $\hat a _l(t)$ is the Heisenberg field operator, 
\begin{align} 
\begin{aligned}
\hat a _l(t) = \hat{\mathcal U}(t)^{\dag}\,\hat a _l\, \hat{\mathcal U}(t), \ \ \ 
\hat{\mathcal U}(t) = \exp (- i \hat{\mathcal H} t )
. 
\end{aligned}%
\end{align}%
It can be also written as an object in the Schr\"odinger picture, 
\begin{align} 
\begin{aligned}
\rho _{lk}(t) = \left\langle \Psi_0(t)\left|\hat a^{\dag}_k \hat a _l\right|\Psi_0(t)\right\rangle, 
\end{aligned}%
\end{align}%
where $\left|\Psi_0(t)\right\rangle = \hat {\mathcal U}(t)\left|\Psi_0\right\rangle$. 
The order or matrix indices was chosen so as to have a 
natural property, 
\begin{align} 
\begin{aligned}
\left\langle \Psi_0(t)\left|\widehat{A}\right|\Psi_0(t)\right\rangle 
= {\rm Tr}\, \rho(t) A ,
\end{aligned}%
\end{align}%
for any single-body operator $A$. 
The ``wide hat'' symbol 
is defined by Eq.~(\ref{eq:ToFock}).

In the mean-field approach, the density matrix is expressed by the 
mean-field amplitude, ${\mbox{\rm\boldmath$\zeta$}}(t)$, 
as: 
\begin{align} 
\begin{aligned}
\rho^{\text{MF}} _{lk}(t) = \zeta_l(t)\zeta_k^*(t) 
. 
\end{aligned}%
\label{eq:MFEigSVD} 
\end{align}%
Mathematically speaking, this relation stipulates that 
$\rho^{\text{MF}}$ has 
a single nonzero eigenvalue equal to $N$, with $N^{-1/2}{\mbox{\rm\boldmath$\zeta$}}(t)$ being the 
corresponding eigenvector. 
It is then only natural to introduce effective 
mean-field amplitudes 
as eigenvectors of $\rho (t)$:
\begin{align} 
\rho _{lk}(t) &= \sum_{s=1,2,3}r_s(t)\,\zeta_l^{(s)}(t)\,\zeta_k^{(s)*}(t) , 
\ \ \ 
{\mbox{\rm\boldmath$\zeta$}}^{(s)}
\cdot
{\mbox{\rm\boldmath$\zeta$}}^{(s')\ast} 
= N\delta_{ss'}
 . 
\label{eq:EigenV} 
\end{align}%
In general, we have three eigenvectors of $\rho(t)$, 
$N^{-1/2}{\mbox{\rm\boldmath$\zeta$}}^{(s)}(t)$, 
and three related eigenvalues,  
$Nr_s(t)$, $r_1(t)+r_2(t)+r_3(t)=1$.  
For a mean-field approximation  
to be recovered, there must 
exist a dominant eigenvalue, $r_1(t) \gg r_2(t),r_3(t)$. 
Below in Section~\ref{SecResDisc} we shall see that it is exactly the situation 
occuring at early stages of the condensate evolution. 
At later stages all three eigenvalues are of the same order of magnitude  
so that the mean-field approximation breaks down completely. 

\section{Time evolution of atomic coherence}
\label{SecCohEv}
\subsection{Irreducible tensor operators}
\label{IrrTOp}
The physical information contained in the correlations expressed by 
the single-body density matrix is {\em atomic coherence\/} \cite{Shore1989}. 
A natural framework for 
discussing such correlations is the picture of irreducible 
tensor operators \cite{Shore1989,Omont1977,Varshalovich1988}. 
An irreducible tensor operator of angular momentum $j$ 
is a set of $2j+1$ square matrices, 
$T_{jm}$, $m=-j,-j+1,\cdots,j$. 
Under three dimensional rotations, they obey the same transfomation laws as 
spherical harmonics $Y_{jm}(\theta,\phi)$, thus forming a natural basis for 
expressing any physical quantity with angular momentum $j$. 
Furthermore, irreducible tensor operators form a full orthonormal set 
in the space of matrices in respect of the Hilbert-Schmidt norm: 
\begin{align} 
\begin{aligned}
{\rm Tr}\,T_{jm}T^{\dagger}_{j'm'} = \delta_{jj'}\delta_{mm'},  
\end{aligned}%
\end{align}%
and for any $(2J+1)\times (2J+1)$ matrix $A$, 
\begin{align} 
\begin{aligned}
A = \sum_{j=0}^{2J}\sum_{m=-j}^jc_{j}^m\,T_{jm}, \ \ \ 
c_{j}^m = {\rm Tr}\,T_{jm}^{\dagger}A
. 
\end{aligned}%
\end{align}%

\subsection{Dipole and quadrupole momenta of the spin-1 condensate}
\subsubsection{Cyclic basis}
For the spin-1 condensate, 
the single-body density matrix (\protect\ref{eq:rhott}) is a 
$3\times 3$ matrix and hence only contains angular 
momenta 0, 1, and 2:
\begin{align} 
\begin{aligned}
\rho(t) = c_0^0 T_{00} \
+ \sum_{m =-1}^1 c_{1 }^m T_{1m }
+ \sum_{m =-2}^2 c_{2 }^m T_{2m } .
\end{aligned}%
\end{align}%
$c_{0}^0$  and 
$c_{1 }^m$ have a very simple interpretation: 
$c_{0}^0$ is in essence the total population, 
\begin{align} 
\begin{aligned}
c_{0}^0 = \frac{1}{\sqrt{3}} \left\langle \hat {\mathcal N}\right\rangle = \frac{N}{\sqrt{3}}\, , 
\end{aligned}%
\end{align}%
and 
$c_{1 }^m$ is related to magnetisation, 
\begin{align} 
\begin{aligned}
c_{1}^m = - \frac{1}{\sqrt{2}} 
\left\langle \hat {\mathcal F}_{m}\right\rangle ^*
, 
\end{aligned}%
\end{align}%
where the spherical components of the angular momentum operator 
are defined as, 
$\hat {\mathcal F}_{\pm} = \mp  \left(
\hat {\mathcal F}_x \pm i \hat {\mathcal F}_y 
\right)/{\sqrt{2}} $, $ 
\hat {\mathcal F}_{0} = \hat {\mathcal F}_z $. 
$c_{2\mu }$ characterise the quadrupole momentum (alignment) 
of the atomic state. 

\subsubsection{Cartesian basis}
It is instructive to rewrite $\rho (t)$ in the 
Cartesian basis, (with $\alpha,\beta = x,y,z$) 
\begin{gather} 
\begin{gathered} 
\rho _{\alpha\beta}(t) = 
\left\langle \Psi_0\left|\hat a^{\dag}_{\beta }(t) \hat a _{\alpha}(t)\right|\Psi_0\right\rangle , 
\\
\hat a_{\pm} = \mp  \left(
\hat a_x \pm i \hat a_y 
\right)/{\sqrt{2}} , \ \ \  
\hat a_{0} = \hat a_z . 
\end{gathered}%
\label{eq:ToCart} 
\end{gather}%
Then, atomic   
population corresponds to the trace of the density 
matrix, 
magnetisation---to its antisymmetric part, and alignment%
---to its traceless symmetric part. 
Taking the symmetric part of an Hermitian matrix is the same as taking 
its real part, hence $\Re\rho $ contains a mix of 
the population and 
alignment 
while $\Im \rho $ is pure magnetisation. 
Explicitly, 
\begin{align} 
\begin{aligned}
\rho _{\alpha \beta } & 
= \Re \rho _{\alpha \beta } 
+ \Im \rho _{\alpha \beta } 
= \Re \rho _{\alpha \beta } 
+ \frac{i}{2} \sum_{\sigma }
\varepsilon _{\alpha\beta \sigma }{\mathcal F}_{\sigma }
, 
\ \ \ 
{\mathcal F}_{\sigma } = -i \sum_{\alpha,\beta }
\varepsilon _{\alpha\beta \sigma }\rho _{\alpha \beta } 
, 
\end{aligned}%
\label{eq:RhoByReF} 
\end{align}%
where ${\mbox{\rm\boldmath$\mathcal F$}}=\left\langle \hat{{\mbox{\rm\boldmath${\mathcal F}$}}}\right\rangle $ is 
the magnetisation vector,
and $\varepsilon _{\alpha\beta \sigma }$ 
is the fully antisymmetric tensor, 
$\varepsilon _{\alpha\beta \sigma }= 
-\varepsilon _{\beta \alpha\sigma }= 
-\varepsilon _{\alpha\sigma \beta } $, 
$\varepsilon _{xyz}=1 
$. 
Introducing the quadrupole tensor as per Ref.\ \cite{Varshalovich1988}, 
\begin{align} 
\begin{aligned}
Q_{\alpha\beta } & = 
\frac{N}{3}  \delta _{\alpha\beta } 
-\Re \rho _{\alpha \beta } 
, 
\end{aligned}%
\end{align}%
we find the irreducible-tensor-operator expansion of $\rho (t)$ as 
\begin{align} 
\begin{aligned}
\rho _{\alpha \beta } & = 
\frac{N}{3}  \delta _{\alpha\beta } 
+ \frac{i}{2} \sum_{\sigma }\varepsilon _{\alpha\beta \sigma }
{\mathcal F}_{\sigma }
- Q_{\alpha\beta } 
. 
\end{aligned}%
\label{eq:DefFQ} 
\end{align}%
\subsection{Atomic coherence under the mean-field approximation}
In the mean-field picture, $\rho _{\alpha \beta} = 
\zeta _{\alpha}\zeta^* _{\beta }$, where the Cartesian components 
of ${\mbox{\rm\boldmath$\zeta $}} = {\mbox{\rm\boldmath$\zeta $}}'+i{\mbox{\rm\boldmath$\zeta $}}''$ are found in full analogy to (\ref{eq:ToCart}). 
The population-and-alignment part of the density matrix is then, 
\begin{align} 
\begin{aligned}
\Re\rho_{\alpha \beta} = 
\zeta' _{\alpha}\zeta '_{\beta } + 
\zeta'' _{\alpha}\zeta'' _{\beta }
. 
\end{aligned}%
\label{eq:ReIm} 
\end{align}%
Magnetisation also follows easily, 
\begin{align} 
\begin{aligned}
{\mbox{\rm\boldmath$\mathcal F$}} = - 2\, [{\mbox{\rm\boldmath$\zeta$}}'\times {\mbox{\rm\boldmath$\zeta$}}'' ]
. 
\end{aligned}%
\label{eq:FMeanF} 
\end{align}%
We can assume that either ${\mbox{\rm\boldmath$\zeta $}}$ is real, ${\mbox{\rm\boldmath$\zeta $}}'' = 0 $, 
or the real and imaginary parts of ${\mbox{\rm\boldmath$\zeta $}}$ are 
orthogonal, ${\mbox{\rm\boldmath$\zeta $}}' \perp {\mbox{\rm\boldmath$\zeta $}}'' $. 
Indeed, since we can always find such $\varphi $ that 
\begin{align} 
\begin{aligned}
\Re{\mbox{\rm\boldmath$\zeta $}} \text{e}^{i\varphi }\cdot \Im{\mbox{\rm\boldmath$\zeta $}} \text{e}^{i\varphi } = 
{\mbox{\rm\boldmath$\zeta $}}' \cdot {\mbox{\rm\boldmath$\zeta $}}'' \cos 2\varphi + 
\frac{{\mbox{\rm\boldmath$\zeta $}}^{\prime 2} - {\mbox{\rm\boldmath$\zeta $}}^{\prime \prime 2}}{2}\sin 2 \varphi = 0 
, 
\end{aligned}%
\label{eq:MakeOrt} 
\end{align}%
this orthogonality can be achieved 
by an overall phase transformation. 
Then, Eq.\ (\ref{eq:ReIm}) explicitly diagonalises $\Re\rho$, 
making it evident that in the mean-field approximation 
$\Re\rho$ always has a 
zero eigenvalue. 
Another two eigenvalues equal ${\mbox{\rm\boldmath$\zeta $}}^{\prime 2}$ 
and ${\mbox{\rm\boldmath$\zeta $}}^{\prime \prime 2}$. 
If ${\mbox{\rm\boldmath$\zeta$}} $ is real, two eigenvalues of $\Re\rho$ are zeros, 
leaving the system in a 
maximal-alignment state along ${\mbox{\rm\boldmath$\zeta $}}$. 
In this case, the magnetisation which is related to $\Im \rho$ vanishes. 
The inverse is also true: 
zero magnetisation is 
characteristic of real ${\mbox{\rm\boldmath$\zeta $}}$. 

An immediate word of caution is necessary here. 
Orthogonality of $\Re{\mbox{\rm\boldmath$\zeta$}}$ and  $\Im{\mbox{\rm\boldmath$\zeta$}}$ is not 
preserved by the Gross-Pitaevskii equation (\ref{eq:GP-Equation}). 
One can rewrite (\ref{eq:GP-Equation}) so as to maintain the said 
orthogonality, 
but the resulting equations are rather bulky and do not seem to allow 
for a new insight; 
worse still, under certain conditions the phase evolution is   
discontinuous. 

\subsection{Equations of motion for the atomic coherence}
In order to gain further insight into the evolution of the atomic coherence, 
consider the equations of motion for the single-body density matrix 
written in terms of the magnetisation and 
quadrupole matrix.  
Direct calculation with Hamiltonian (\ref{eq:HFockSpace}) yields, 
\begin{align} 
\begin{aligned}
\dot {\mathcal F}_{\alpha} &= \sum_{\sigma } 
\varepsilon _{\alpha \sigma z}\left(p{\mathcal F}_{\sigma } + 2 q 
Q_{\sigma z}\right) , 
\\ 
\dot Q_{\alpha \beta } &= \sum_{\sigma } 
\varepsilon _{\alpha \sigma z}\left(
p Q_{\sigma \beta } + \frac{q}{2} \delta_{\beta z}{\mathcal F}_{\sigma } 
\right) 
+ \left\{
\alpha \leftrightarrow\beta 
\right\} 
\\ & \hspace{3em} + \left[
\dot Q_{\alpha \beta } 
\right]_{\text{coll.}} 
. 
\end{aligned}%
\label{eq:EvolFQ} 
\end{align}%
The benefits of the representation of irreducible tensor operators are now 
clearly seen. 
Collisions are spherically symmetric so that only the linear and quadratic 
Zeeman effects may couple magnetisation and alignment. 
In actuality, additional reflection symmetries leave only quadratic 
Zeeman effect to do it. 
The linear Zeeman effect leads only to the overall Larmor precession 
around the $z$ axis. 
Furthermore, 
collisions only contribute directly to the evolution 
of the quadrupole, 
\begin{align} 
\begin{aligned}
\left[
\dot Q_{\alpha \beta } 
\right]_{\text{coll.}} = 2 g_2 \Im \sum_{\sigma } \left\langle 
\hat a^{\dag}_{\alpha }
\hat a^{\dag}_{\beta }
\hat a^{2}_{\sigma }
\right\rangle
. 
\end{aligned}%
\end{align}%
In general, this contribution contains nontrivial two-body correlations and 
cannot be simplified any further. 
Within the mean-field approximation, 
\begin{align} 
\begin{aligned}
\big[
\dot Q_{\alpha \beta } 
\big]_{\text{coll.}} = 2g_2\Big[
2\big(
{\mbox{\rm\boldmath$\zeta $}}^{\prime}
\cdot 
{\mbox{\rm\boldmath$\zeta $}}^{\prime \prime}
\big) 
\big(
\zeta '_{\alpha }
\zeta '_{\beta }
- 
\zeta ''_{\alpha }
\zeta ''_{\beta }
\big) 
 - \big(
{\mbox{\rm\boldmath$\zeta $}}^{\prime 2} - {\mbox{\rm\boldmath$\zeta $}}^{\prime \prime 2}
\big) 
 \big(
\zeta '_{\alpha }
\zeta ''_{\beta }
+ 
\zeta ''_{\alpha }
\zeta '_{\beta }
\big) 
\Big]
. 
\end{aligned}%
\label{eq:QCollMF} 
\end{align}%
Note that here we do not assume that  
${\mbox{\rm\boldmath$\zeta $}}' \perp {\mbox{\rm\boldmath$\zeta $}}'' $.

Equations (\ref{eq:EvolFQ}) allow one to make 
qualitative statements about the evolution of the system's 
coherence independently of the details of 
collisional interactions. 
Assume that the initial state of the system is 
as in experiment \cite{Sengstock2004}. 
Then, the initial magnetization vector is along the $x$ axis, 
${\mathcal F}_z=0$. Since $\dot{\mathcal F}_z=0$, 
cf.~Eqs.\ (\protect\ref{eq:EvolFQ}), 
magnetization ${\mbox{\rm\boldmath$\mathcal F$}}$ never leaves the $xy$ plane. 
As a result, 
under the mean field 
approximation the alignment eigenvectors stay in a plane 
orthogonal to the $xy$ plane, cf.~(\ref{eq:FMeanF}). 
If at a particular time magnetisation vanishes, the mean field 
picture predicts that the system must be axially symmetric. 
Hence, under the SMA, absence of axial symmetry at zero 
magnetisation is 
a clear signature of many-body effects. 
Since thermalisation via symmetric collisions 
can only increase the symmetry 
of the system, 
it appears plausible that this statement should survive 
beyond the SMA. 
Verifying this conjecture is subject to further analysis. 
\section{Results and discussion}\label{SecResDisc}
\subsection{Fock space of the spinor condensate}
The choice of the basis in the Fock space is inherently 
linked to the angular-momentum properties of Hamiltonian (\ref{eq:HFockSpace}). 
A native 
bookkeeping is to introduce the basis 
states by the populations 
of the single-particle states with $F_z = 1,0,-1$. 
By construction, such basis state in the Fock space 
is an eigenstate of the three mode population 
operators: 
\begin{align} 
\begin{aligned}
\hat{\mathcal N}_k \left|N_1,N_0,N_{-1}\right\rangle = 
N_k \left|N_1,N_0,N_{-1}\right\rangle ,\ \ \ k=1,0,-1. 
\end{aligned}%
\end{align}%
These states also happen to be the 
eigenstates of $\hat {\mathcal N}$ and $\hat {\mathcal F}_z$: 
\begin{align} 
\begin{aligned}
\hat {\mathcal N}\left|N_1,N_0,N_{-1}\right\rangle &= 
N \left|N_1,N_0,N_{-1}\right\rangle,  \\
\hat {\mathcal F}_z\left|N_1,N_0,N_{-1}\right\rangle &= 
M\left|N_1,N_0,N_{-1}\right\rangle 
, 
\end{aligned}%
\end{align}%
where $N=N_1+N_0+N_{-1}$ and $M = N_1-N_{-1}$. 
Since $\hat {\mathcal H}$ commutes with $\hat {\mathcal N}$ and 
$\hat {\mathcal F}_z$, states with different $N$'s ans
$M$'s are not mixed by free evolution. Another simplification 
when using this ``natural'' basis 
is the ease with which the many-body operators 
are implemented (Appendix \ref{ToFock}). 

In the absence of the quadratic Zeeman effect (QZE), Hamiltonian (\ref{eq:HFockSpace}) conserves the full angular momentum, 
$\hat{{\mbox{\rm\boldmath${\mathcal F}$}}}^2$, and its component, 
$\hat {\mathcal F}_z $. 
In this case the time evolution is solved 
\cite{Koashi2000} by simply classifying the basis 
states by 
the eigenvalues of the three operators: 
\begin{align} 
\begin{aligned}
\hat{{{\mathcal N}}}\left|N,F,M\right\rangle 
_{\mathcal F}
&= N\left|N,F,M\right\rangle 
_{\mathcal F}, \\ \hat{{\mbox{\rm\boldmath${\mathcal F}$}}}^2\left|N,F,M\right\rangle _{\mathcal F}
&= F(F+1)\left|N,F,M\right\rangle 
_{\mathcal F}, \\
\hat {\mathcal F}_z\left|N,F,M\right\rangle _{\mathcal F}
&= M\left|N,F,M\right\rangle _{\mathcal F} . 
\end{aligned}%
\end{align}%
Here, 
$0\leq F\leq N$ and $-F\leq M\leq F$. 
The subscript ${\mathcal F}$ distinguishes the ``angular momentum'' basis 
from the ``natural'' basis introduced above. 

In the presence of the QZE, 
$F$ is not a good quantum number any more, while $M$ is still conserved. 
Hamiltonian (\ref{eq:HFockSpace}) can then be diagonalised within subspaces 
of given $M$ of the ``natural'' basis. 
Conversely, rotating the states by the radio-frequency field as in Ref.\ \cite{Sengstock2004} conserves 
$F$ while mixing different $M$'s. 
Using the ``angular momentum'' basis may thus be of some help. 
In the calculations summarised in this paper, the natural basis was used. 
Exploring the advantages of the angular momentum basis remains a subject to 
further work. 
\subsection{Numerical procedure and the choice of parameters}
All simulations summarised in this paper were 
carried out with the experiment 
by Sengstock and co-workers in mind \cite{Sengstock2004}. 
In this experiment, the condensate is firstly prepared in 
a state where all atomic spins are antiparallel 
to the $z$ axis. 
A radio-frequency pulse is then used to re-orient the spins antiparallel 
to the $x$ axis. 
In our notation, this state looks as
\begin{align} 
\begin{aligned}
\left|\Psi_0\right\rangle = \exp \left(
-i \vartheta \hat {\mathcal F}_y
\right) \left|0,0,N\right\rangle
, 
\end{aligned}%
\label{eq:InitPsi} 
\end{align}%
where $\vartheta =\pi /2$. 
Since the manifold of $M = -N$ is non-degenerate, 
$\left|0,0,N\right\rangle$ is also an eigenstate of the total angular 
momentum,  
$\left|0,0,N\right\rangle = \left|N,N,-N\right\rangle_{\mathcal F}$. Furthermore, since the evolution does 
not mix different $M$'s, 
$\left|0,0,N\right\rangle$ must be an eigenstate of the 
Hamiltonian, providing us with a test of numerics. 
Another test is setting 
$\vartheta = \pi$, yielding $\left|\Psi_0\right\rangle = \left|N,0,0\right\rangle = \left|N,N,N\right\rangle_{\mathcal F}$, 
which is also an eingenstate of the Hamiltonian. 
Having thus checked that both for $\vartheta = 0$ 
and for $\vartheta = \pi$ ``nothing 
at all happens,'' we 
proceeded with a series of simulations with $\vartheta = \pi/2$. 

To compare the many-body results to the mean-field approximation, we also 
ran a mean-field simulation of the system using (\ref{eq:GP-Equation}). 
It easy to see that, with our choice of the many-body initial state, 
Eqs.\ (\protect\ref{eq:MFEigSVD}) are exact at $t=0$. 
This provides us with the initial condition for the mean-field 
simulation. 
This is certainly the same initial mean-field amplitide as used in 
Refs. \cite{Sengstock2004,Sengstock2005}.

Simulations presented in this paper were run for the following set 
of parameters: $p=0$, $q=1$, and $Ng_2 = - 0.3$. 
These differ from the parameters characteristic of 
experiments 
\cite{Sengstock2004,Sengstock2005}, $p\gg q$, $Ng_2 = -0.03q$, 
by setting $p$ to zero, and enhancing nonlinearity by 
an order of magnitude. 
It should be emphasised that 
observation times in these experiments are too low for the many-body effects to manifest themselves. 
Simply extending the ``observation'' time in 
simulations allows for the many body effects to 
be seen, but produces two many oscillations of physical 
quantities on the relevant time scales. 
Enhancing the nonlinearity allows one to produce comprehensible graphics, 
while making no qualitative difference to the condensate evolution. 
Setting $p$ to zero is of no consequence: in essence this means introducing 
a reference frame following the Larmor precession. 

The number of particles in our simulations was $N=50$. 
Importantly, this number 
is large enough to clearly separate the relaxation and the 
revival time scales. 
Indeed, relaxation is exacty that what we expect from a many body 
treatment, whereas revivals are artifact of the model. 
At the same time, leaving the atom number relatively low 
simplifies numerics enourmously, making the problem easily 
tractable on an 
average desktop computer. 
\begin{figure}[h]
 \begin{center} 
 \includegraphics[width=0.5\textwidth]{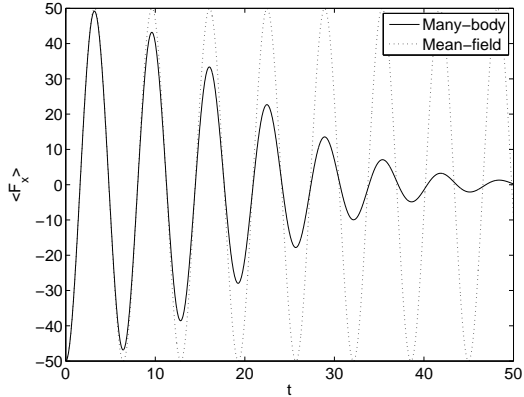}
 \caption{Evolution of the condensate 
	 magnetisation as function of the scaled time. 
	 $N=50$, $p=0$, $q=1$, $Ng_2 = -0.3$. 
	 Only the $x$ component of the magnetisation is nonzero. 
	 In the mean-field approximation, magnetisation oscillates without relaxation.}
\label{fig:F}
\end{center}
\end{figure}
\begin{figure}[h]
 \begin{center} 
 \includegraphics[width=0.5\textwidth]{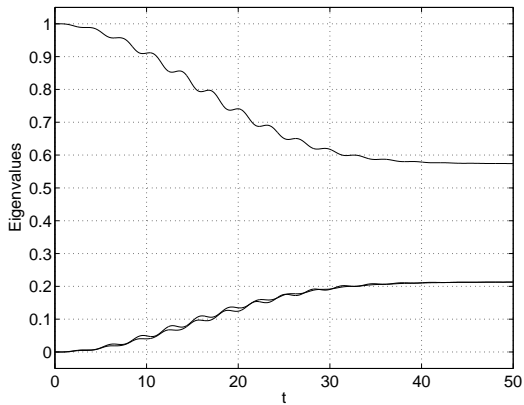}
 \caption{Scaled eigenvalues of the single-time 
	 density matrix. $N=50$, $p=0$, $q=1$, $Ng_2 = -0.3$. 
	 In the mean field approximation, one eigenvalue always equals 
	 one and the other two eingenvalues are
	 zero.}
\label{fig:eigen}
\end{center}
\end{figure}
\subsection{Evolution of atomic coherence}
\label{SecNumCoher}
Figure \ref{fig:F} 
shows the evolution of the condensate magnetisation, 
Eq.\ (\protect\ref{eq:FMeanF}), 
calculated both within the many-body model and under the mean-field 
approximation. 
With our choice of the inital state and parameters, only the $x$ 
component of the magnetisation is nonzero. 
In the mean-field approximation, this component oscillates 
periodically between values $-N$ and $N$. 
{\em Exact\/} periodicity follows from the analytical solution 
\cite{Sengstock2005}. 
The many body result shows {\em decaying\/} 
oscillations with the same period. 
Qualitatively, the decay can be understood if 
considering the simplest generic model of collisional nonlinearity: 
the Kerr oscillator \cite{Kerr}. 
The latter is a single-mode quantum oscillator with quartic nonlinearity, 
\begin{align} 
\begin{aligned}
{\mathcal H}_{\text{eff}} = \omega_{\text{eff}} \hat n 
+ \frac{g_{\text{eff}} }{2} \hat n (\hat n -1). 
\end{aligned}%
\end{align}%
Assume the Kerr oscillator is 
initially in a coherent state with amplitude $\alpha$. 
A straightforward 
calculation of the average position of the oscilator yields, 
\begin{multline} 
\left\langle \hat x(t) \right\rangle = 
\frac{1}{\sqrt{2}} \Re\big\{ \alpha \exp 
\big[
-i\omega_{\text{eff}} t + N_{\text{eff}}
\big({\rm e}^{-ig_{\text{eff}}t} - 1\big) 
\big]\big\} 
\\
\approx x_{\text{MF}}(t)\,{\rm e}^{-N_{\text{eff}}g^2_{\text{eff}}t^2/2}, 
\ \ \ |g_{\text{eff}}t| \ll 1  . 
\end{multline}%
where $N_{\text{eff}}=|\alpha|^2$ and 
\begin{align} 
\begin{aligned}
x_{\text{MF}}(t) = \frac{1}{\sqrt{2}} \Re\big[ \alpha {\rm e}^{ 
-i(\omega_{\text{eff}} + g_{\text{eff}}N_{\text{eff}})t
}\big] 
\end{aligned}%
\end{align}%
is the mean-field approximation to $\left\langle \hat x(t) \right\rangle$ 
which accounts for the frequency shift but not the dephasing. 
Given the obvious crudeness of the Kerr model, 
it reproduces the evolution of the magnetisation, Fig.~\ref{fig:F}, with 
surprising accuracy. 
Setting $N_{\text{eff}}g^2_{\text{eff}} \approx 2 N g^2_2 $ 
results in a good 
quantitative estimate of the decay time, 
$\tau _{\text{decay}} \approx (\sqrt{N}g_2)^{-1}\approx 23 $. 
The Gaussian nature of the decay is also evident in Figs.~\ref{fig:F} 
and \ref{fig:eigen}.

Contrary to what could be expected, 
the system does not evolve towards a spherically symmetric state. 
A formal indication of this may be seen in 
Fig.\ \ref{fig:eigen}. This figure 
shows scaled eigenvalues of the single-body density matrix, 
cf.\ Eq.\ (\protect\ref{eq:EigenV}). 
In the mean field approximation, one eigenvalue always equals 
one and the other two eingenvalues are
zero. 
This holds for the initial state, but for nonzero times we find 
that all 
three eigenvalues are nonzero. 
Most interestingly, they do not evolve towards the value of 1/3 
indicative of spherical symmetry. 
Two of them stabilise at $\approx 0.22$, while the third one 
remains close to $\approx 0.56$. 
Since the magnetisation decays completely, cf.\ Fig.\ \ref{fig:F}, 
the persistent asymmetry must be of quadrupole nature. 
\begin{figure}[h]
 \begin{center} 
 \includegraphics[width=0.49\textwidth]{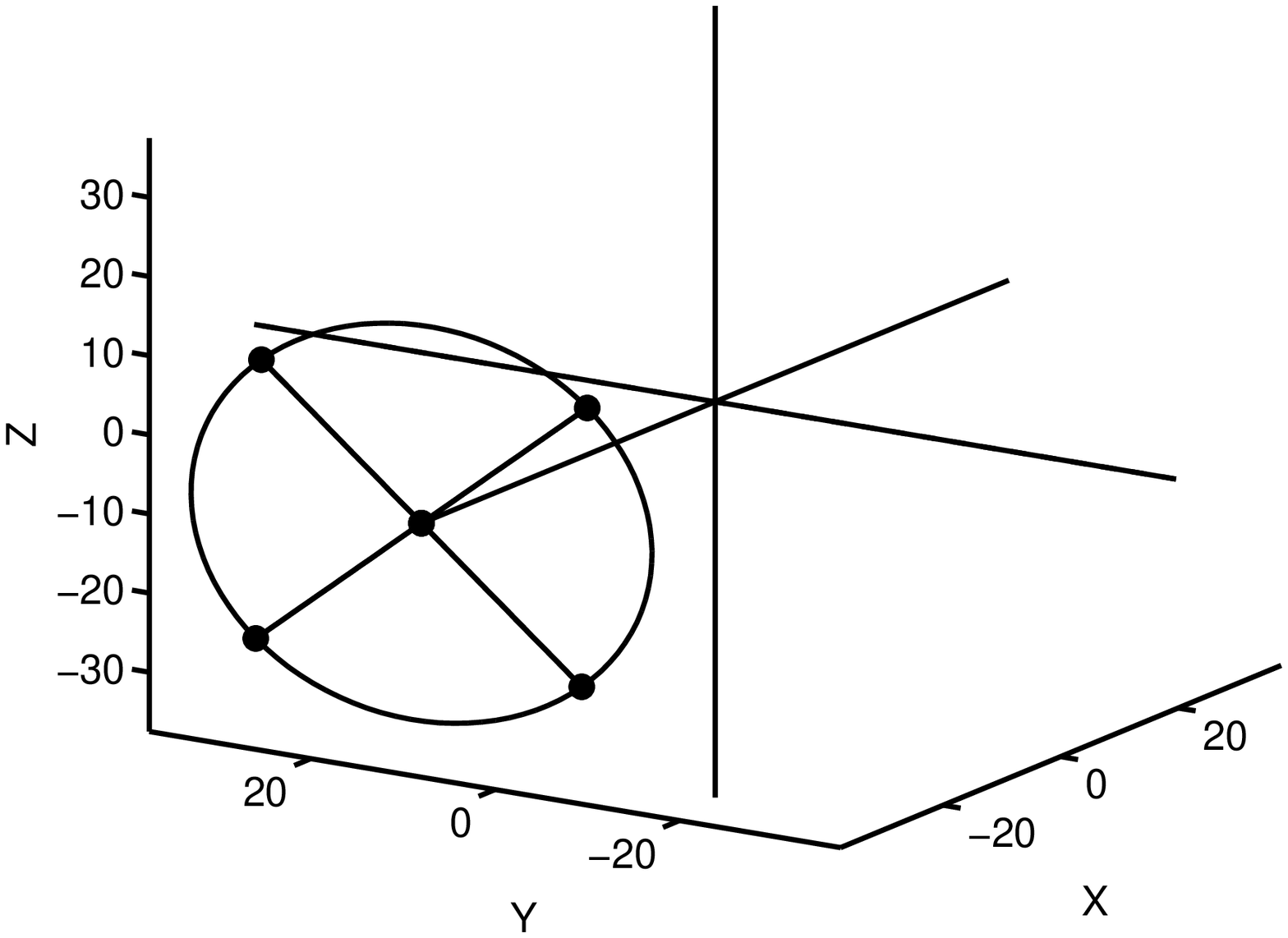}
 \includegraphics[width=0.49\textwidth]{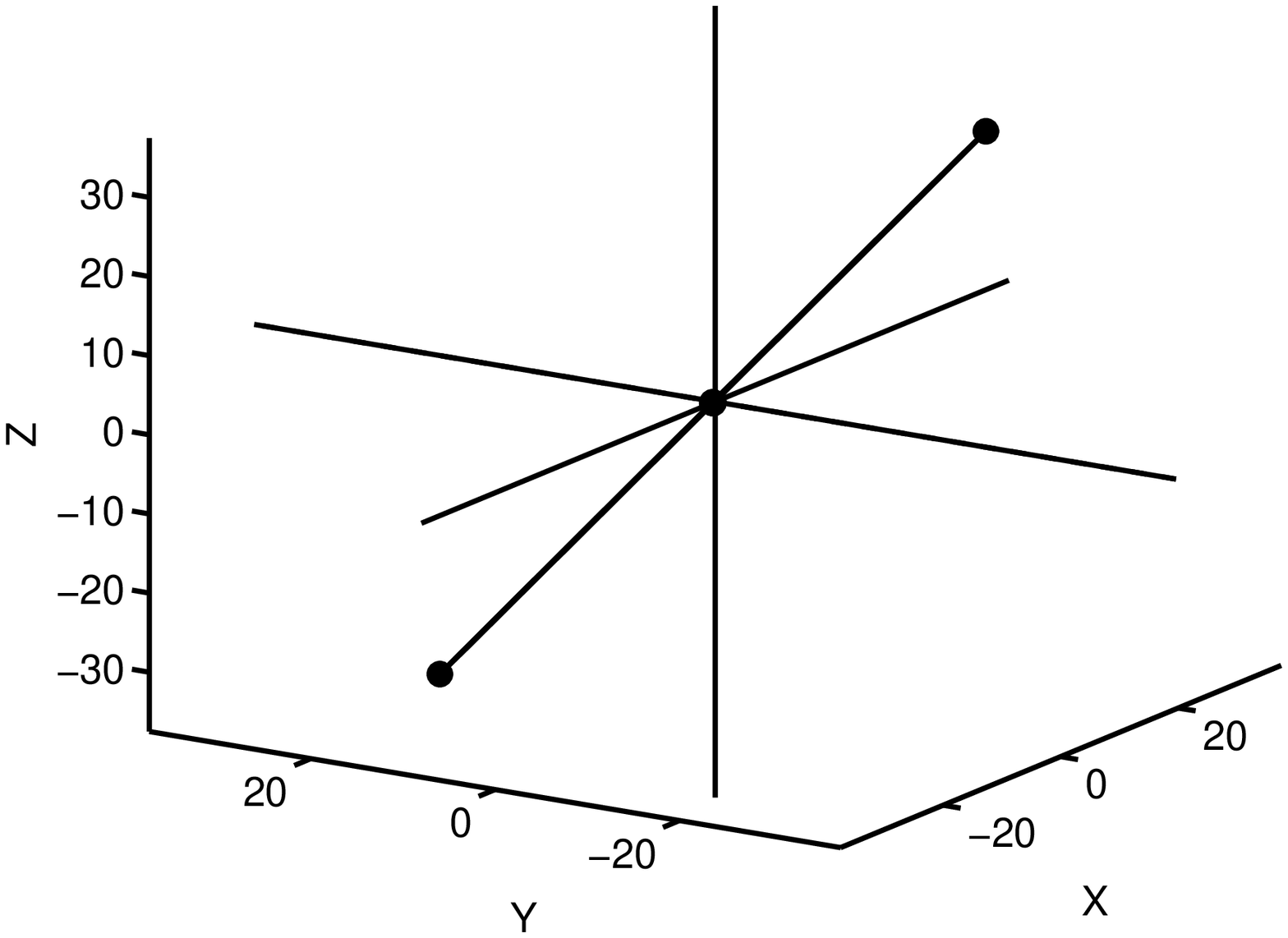}
 \includegraphics[width=0.49\textwidth]{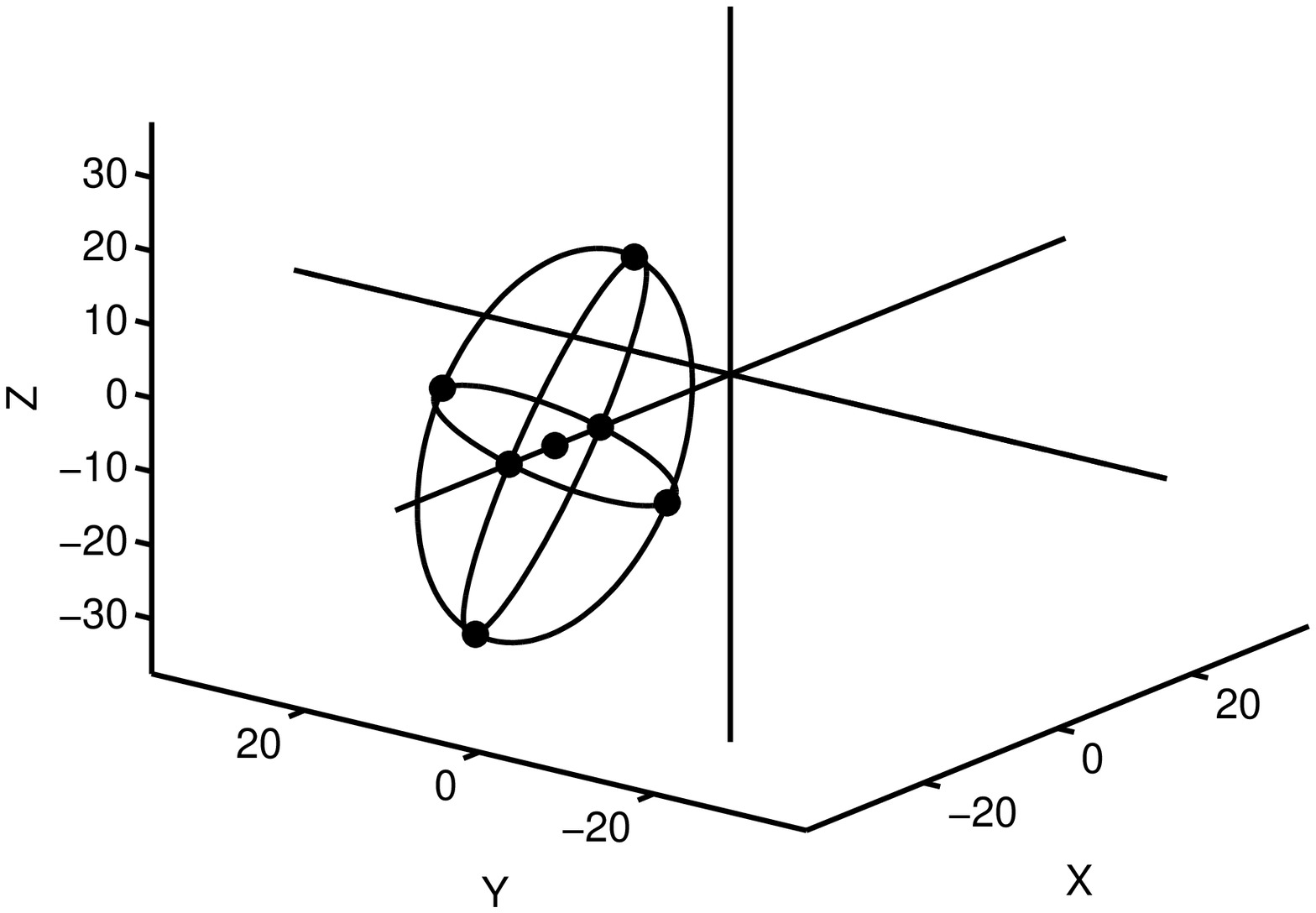}
 \includegraphics[width=0.49\textwidth]{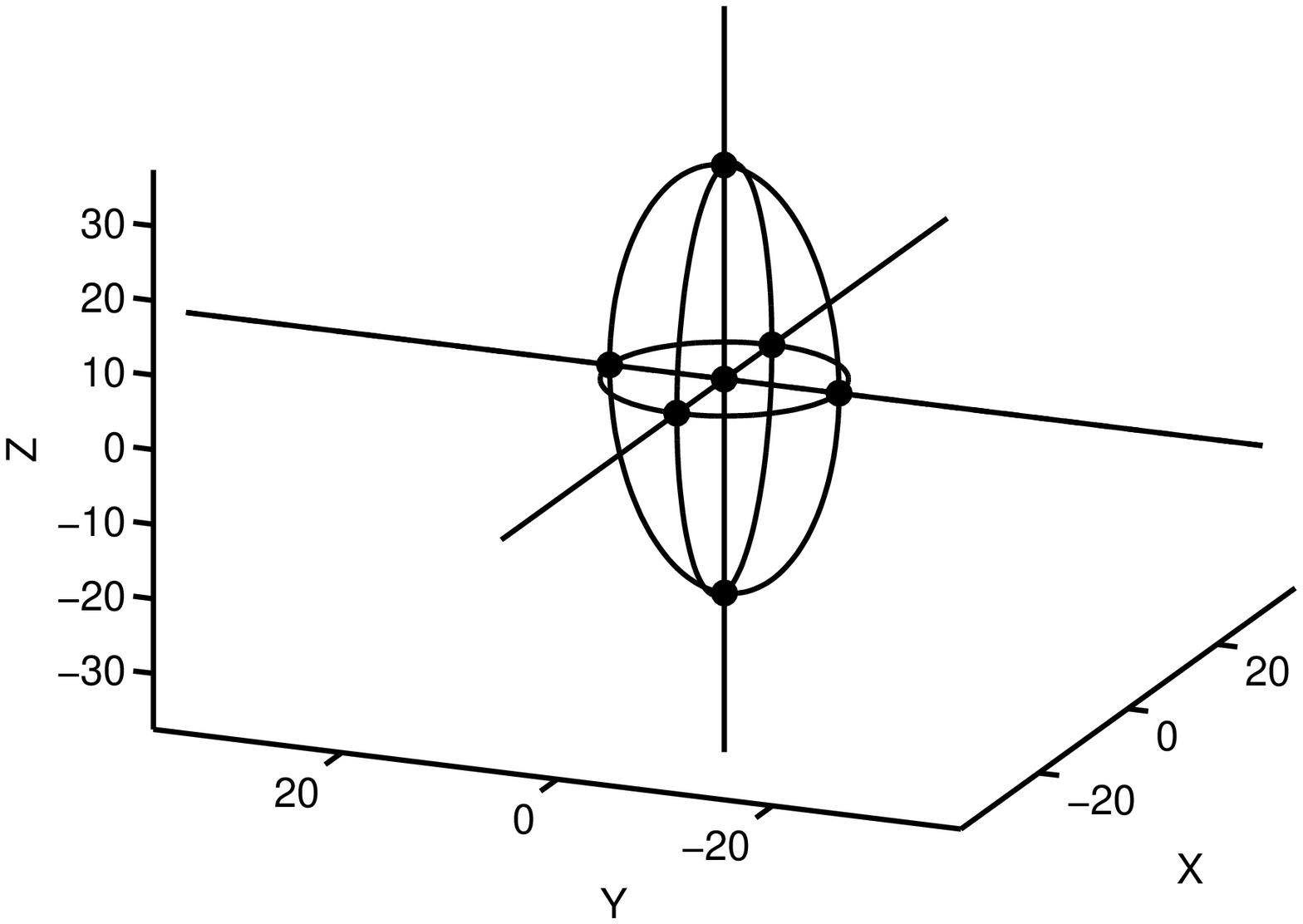}
 \caption{Visualisation of the time evolution of the 
	 condensate coherence. The pictures 
	 shows the alignment ellipsoid, the centre of which is shifted by 
	 the magnetisation vector. One of the alignment axis 
	 is always collinear with the magnetisation vector, so that the 
	 only parameter specifying the ellipsoid's orientation is the tilt angle. 
	 In the mean-field approximation, the ellisoid reduces to a flat 
	 ellips orthogonal to the magnetisation vector. 
	 {\em Top row:\/} initial stages of the evolution 
	 at which the mean-field picture is 
	 valid. 
	 {\em Left:\/} the initial state of the  
	 condensate coherence. The alignment ellipsoid is reduced to 
	 a disc. 
	 {\em Right:\/} the condensate coherence at zero magnetisation. 
	 The alignment ellipsoid is reduced to a straght line in the $yz$ plane with $\pi/4$ tilt. 
	 {\em Bottom row:\/} later stages of the 
	 evolution at which the mean-field picture 
	 fails. 
	 {\em Left:\/} a typical picture of coherence at intermediate stages of the 
	 evolution. 
	 {\em Right:\/} the final state of 
	 condensate coherence. The magnetisation and tilt are zero, and 
	 the 
	 ellipsoid is axially symmetric in respect of the quantisation axis.}
\label{fig:visual}
\end{center}
\end{figure}

\begin{figure}[h]
 \begin{center} 
 \includegraphics[width=0.49\textwidth]{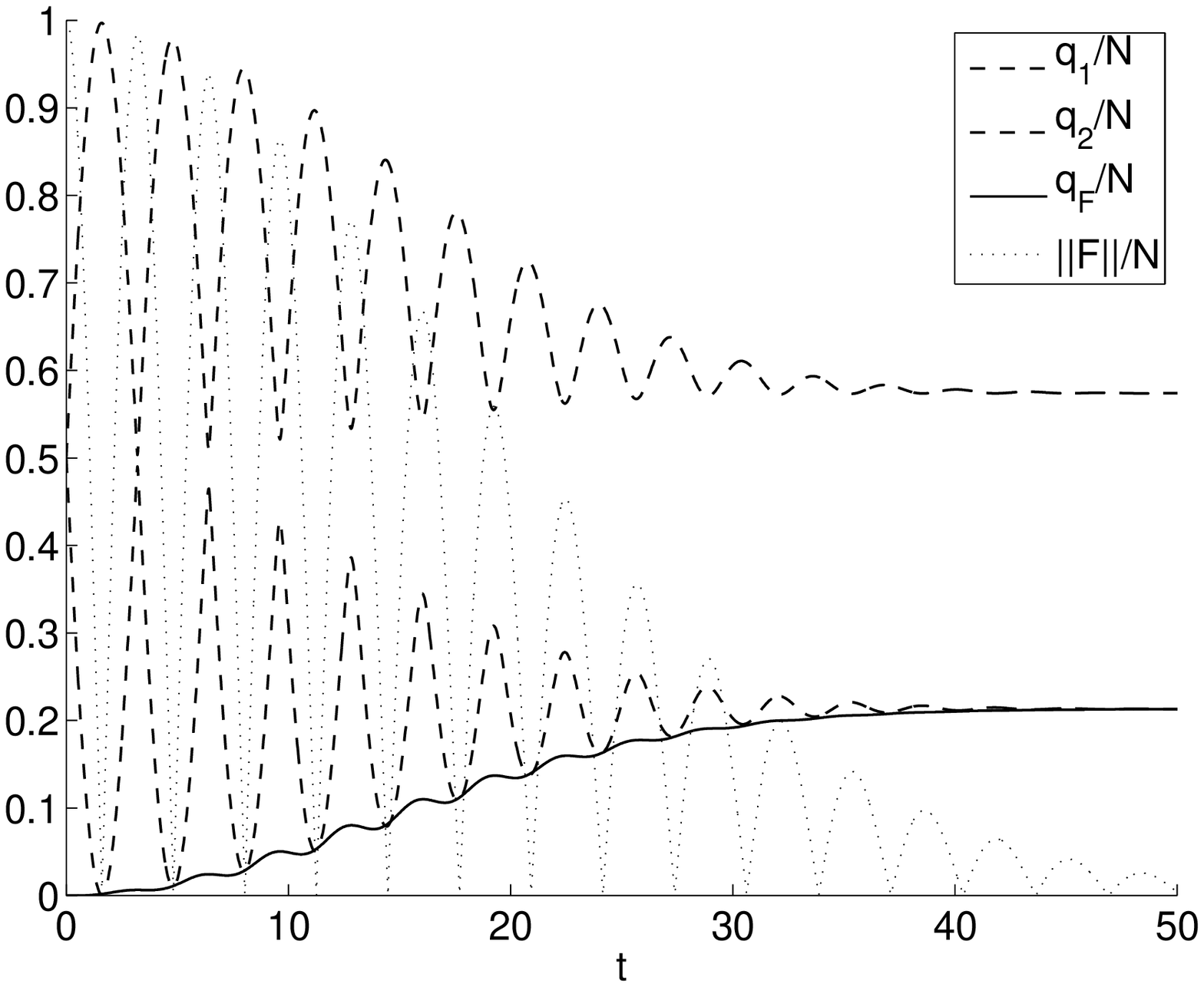}
 \includegraphics[width=0.49\textwidth]{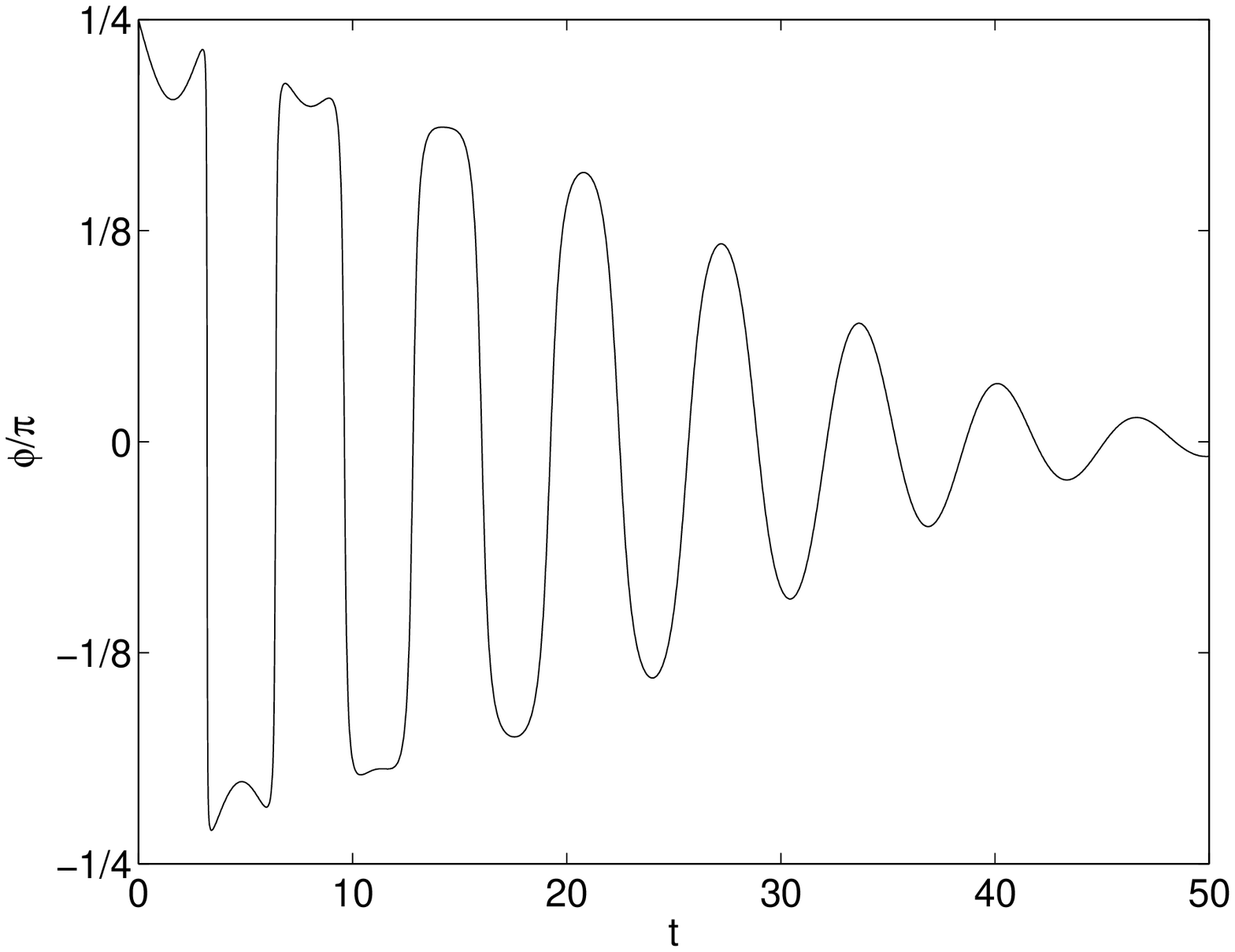}
 \caption{Evolution of the quadrupole. $N=50$, $p=0$, $q=1$, $Ng_2 = -0.3$. Left: the alignment eigenvalues, right: the tilt angle. 
	 The smallest eigenvalue corresponding to the alignment axis along 
	 the magnetisation  
	 is shown 
	 as solid line, the other two---as dashed lines. 
	 The modulus of magnetisation 
	 (dotted line) helps to visually synchronise this picture with Fig.\ \ref{fig:F}. 
	 In the mean field approximation, the smallest eigenvalue is exactly zero  
	 while the other two 
	 eigenvalues and the tilt oscillate periodically without relaxation.}
\label{fig:Q}
\end{center}
\end{figure}

Being 
a mathematically rigorous way of characterising alignment, 
the quadrupole tensor is 
not the best choice so far as {\em visualisation\/} of 
the results is concerned. 
A much more convenient way is thinking in terms of 
the {\em alignment ellipsoid\/}. 
The latter is found by diagonalising the real part of the 
density matrix, cf.\ Eq.~(\ref{eq:RhoByReF}). 
The eigenvectors of $\Re\rho $ are the 
alignment axes. 
The eigenvalues of $\Re\rho $ are positive, 
their sum equals population, 
$q_1+q_2+q_3=N$, and 
differences between them characterise alignment. 
In particular, three equal eigenvalues, $q_1=q_2=q_3=N/3$, 
mean the atomic state has no alignment at all. 
The system of three orthogonal eigenvectors and three positive 
eigenvalues is naturally  
visualised as a three-dimensional ellipsoid. 
If we additionally shift the centre of this 
alignment ellipsoid by the magnetisation vector, 
the resulting picture contains all details of the atomic coherence, 
cf.\  Figure \ref{fig:visual}. 
An additional simplification is that the smallest axis of the ellipsoid 
was always found to be collinear with the magnetisation vector. 
This also holds if $p\neq 0$ when the whole picture is subject to  
the Larmor precession.  
Thus the only parameter we need in order to fully specify the orientation 
of the alignment ellipsoid is the tilt angle; the latter is defined 
as the angle between the largest axis of the ellipsoid and the 
quantisation ($z$) axis. 
Together with the eigenvalues of $\Re \rho$ 
[cf.\ Eq.\ (\protect\ref{eq:RhoByReF})] and the 
magnetisation vector, 
this angle completes characterisation of the atomic coherence. 

Visualisation of atomic coherence at 
different stages of the evolution may be seen in 
Fig.\ \ref{fig:visual}. 
Evolution of the alignment eigenvalues and of the tilt angle is shown in 
Fig.\ \ref{fig:Q}. 
At early stages of the evolution, the mean field picture holds with 
a very good accuracy. 
In the mean field approximation, the alignment ellipsoid reduces to 
a flat ellips orthogonal to the magnetisation vector. 
There are two characteristic moments 
of the mean-field evolution [cf.~Eq.\ (\protect\ref{eq:FMeanF}) and Fig.\ \ref{fig:Q}, top row]. 
If the magnetisation is maximal, the alignment axis are equal; 
the alignment ellipsoid becomes a disk 
(Fig.\ \ref{fig:Q}, top-left). 
If the magnetisation is zero, only one alignment eigenvalue is nonzero; 
the alignment ellipsoid reduces to a straight line at $\pi/4$ tilt 
(Fig.\ \ref{fig:Q}, top-right). 
The mean-field evolution of the atomic coherence is in essence 
periodic oscillations between these charateristic configurations. 

On later stages of the evolution the mean field picture breaks down. 
All three eigenvalues of  $\Re\rho $ are nonzero, so that 
the alignment ellipsoid acquires thickness, 
becoming an ellipsoid proper (Fig.\ \ref{fig:visual}, bottom-left). 
The tilt oscillations are damped and finally cease 
(Fig.\ \ref{fig:Q}). 
The final state of atomic coherence is depicted in Fig.\ \ref{fig:visual}, 
bottom-right. 
The magnetisation is zero so that the ellipsoid is centered at the centre of 
coordinates. The larger axis is along $z$, the other two are equal 
so that the alignment ellipsoid is axially symmetric. 
This state of atomic symmetry is impossible to obtain in the mean-field 
picture. 

Thus in the course of the evolution atomic coherence 
undergoes a complete 
change of symmetry. Initially, the system is prepared 
in an axially symmetric state 
in respect of the $x$ axis. At intermediate stages of the evolution, all 
symmetries are broken. Finally, the system again tends towards 
an axial symmetry. 
Not unexpectedly, the final symmetry coinsides 
with the dynamical symmetry, with 
the symmetry axis parallel to the magnetic field. 

\section{Conclusion}
We have modelled the quantum many-body dynamics of a spin-1 condensate 
numerically within a single mode approximation, 
using the representation of irreducible tensor operators so as 
to conveniently visualise the condensate evolution. 
Comparing results of the numerical simulation to those obtained 
within the mean field approximation shows a number of characteristic 
features of the many-body evolution which are impossible to obtain 
under the mean field approximation. 
In particular, the final state of the condensate coherence is 
of a characteristic 
quadrupole nature, axially symmetric in respect of the 
quantisation axes, whereas this kind of atomic coherence is impossible 
in principle in the mean field approximation.  
Experimentally observing this state of atomic coherence 
would be a distinct signature of spin-mixing collisions without 
the mean-field approximation. 
\section*{Acknowledgments} L.P.\ and R.W.\ acknowledge 
numerous stimulating discussions with K.\ Bongs, K.\ Sengstock, 
J.\ Kronj\"ager and P.\ Navez. 
L.P.\ gratefully acknowledges support of the 
Program Atomoptik of the Landesstiftung Baden-W\"urtemberg. 

\appendix
\section{Implementation of many-body operators in the Fock space.}
\label{ToFock}
To be specific, consider how mode annihilation operators, $\hat a_k$, are 
implemented in a 
high-level numerical package. 
Such packages 
handle sparse rectangular matrices very effectively, so the first step 
is to replace a triple index, $\{N_1,N_0,N_{-1}\}$, characterising 
a basis state in the Fock space, by a linear index. A helpful trick is to 
start from introducing an extended index, $n = 1 + N_1 + (N+1) N_0$, where 
$0\leq N_1,N_0\leq N$. 
The third index, $N_{-1}$, is found from 
$N_1 + N_0 + N_{-1} = N$. 
Each value of the extended index, 
$1\leq n\leq (N+1)^2$, thus corresponds to a unique triple, 
\begin{align} 
\begin{aligned}
N_1(n) & = (n-1)\hspace{-0.75em}\mod{(N+1)} , \\
N_0(n) & = \frac{n-1-N_1(n)}{N+1} , \\ 
N_{-1}(n) &= N -N_1(n) - N_0(n) 
. 
\end{aligned}%
\end{align}%
This includes nonphysical entries with $N_{-1}(n)<0$ which are 
to be filtered out later. 
The advantage of using the extended indexing is that, 
as a rule, many-body operators 
written in the extended index are simple banded matrices. For example, 
if $k=1,0$, 
\begin{align} 
\begin{aligned}
\left\langle n'\left|\hat a_k\right|n\right\rangle = \delta_{n',n-\Delta_k}
\sqrt{N_k(n)}, \ \ \Delta_k = (N+1)^{1-k}
. 
\end{aligned}%
\label{eq:AMatr} 
\end{align}%
That is, in the extended index, $\hat a_1$ and $\hat a_0$ are sparse 
matrices with a single nonzero side diagonal. 
Generating such matrices numerically is straightforward;  
$\hat a_{-1}$ which is not found in such simple way can then be obtained 
by a suitable permutation of rows and columns of $\hat a_1$ (say). 
In the matrices generated using (\ref{eq:AMatr}) 
one should retain only columns for 
which $N_{-1}(n)\geq 0$ and rows for which $0\leq N_{-1}(n')\leq N-1$ 
and $ 
N_{1}(n'),N_{0}(n')\leq N-1$. (Recall that $\hat a_k$ acts from the 
subspace of $N$ atoms into the subspace of $N-1$ atoms.) 
This is equivalent to building a 
``physical'' linear index, changing $N_1$ 
first, and $N_0$---second, while 
watching for the condition $N_1+N_0\leq N$; 
however the property of having a single nonzero 
side diagonal does not hold for the 
physical indexing. 
Generalisation to higher-order spin values is straightforward. 

There is a further subtlety in how products of field operators are 
implemented numerically. 
An unwanted by-product of introducing field operators is a formal 
non-conservation of the particle number. 
We wish to restrict ourselves 
to a subspace of given $N$ and introduce annihilation operators 
as acting from this 
subspace into the subspace of $N-1$ atoms. 
Then, only bilinear normally ordered operator products, 
$\hat a_k^{\dag} \hat a_l$, are immediately defined for numerical 
purposes. 
Viewed literally, quartic normally ordered operators products [cf.~the first 
line of (\ref{eq:HFockSpace})] imply knowledge of the field operators 
acting from the subspace of $N-1$ atoms into the subspace of $N-2$ atoms 
and hence cannot be implemented under this restriction. 
The solution to this problem is obvious: use commutation relations. 
So, with ${\mathcal I}_N$ being the unit matrix in the subspace of $N$ atoms,
\begin{gather} 
\begin{gathered} 
\hat a_l \hat a_k^{\dag} = \hat a_k^{\dag} \hat a_l, \ \ k\neq l , 
\\ 
\hat a_k \hat a_k^{\dag} = \hat a_k^{\dag} \hat a_k + {\mathcal I}_N , 
\\ 
\hat a_k^{\dag}\hat a_k^{\dag}\hat a_k\hat a_k = 
(\hat a_k^{\dag}\hat a_k)^2 - \hat a_k^{\dag}\hat a_k, 
\end{gathered}%
\label{eq:CommNum} 
\end{gather} 
etc. [cf. also the second 
line of (\ref{eq:HFockSpace})]. 
Equations (\ref{eq:CommNum}) hold in the complete Fock space, 
but only their right-hand sides 
make numerical sense under the above restriction. 
 

\end{document}